\documentclass[smallextended]{svjour3}       
\smartqed  
\usepackage{graphicx}
\usepackage{amsmath}
\usepackage{amssymb}
\usepackage{graphicx}
\usepackage{color}
\usepackage{bm}
\usepackage{fix-cm}

\usepackage[compress]{natbib}
\bibpunct{[}{]}{,}{n}{,}{,} 

\usepackage[mathscr,scaled=1.15]{urwchancal}
\DeclareFontFamily{OT1}{pzc}{}
\DeclareFontShape{OT1}{pzc}{m}{it}%
{<-> s * [1.15] pzcmi7t}{}
\DeclareMathAlphabet{\mathpzc}{OT1}{pzc}{m}{it}

\journalname{Few-Body Systems}

\begin{document}

\title{Structure of the nucleon and its first radial excitation
%
}


\author{Jorge Segovia}


\institute{Jorge Segovia \at
           Departamento de Sistemas F\'isicos, Qu\'imicos y Naturales, \\
           Universidad Pablo de Olavide, E-41013 Sevilla, Spain \\           
           \email{jsegovia@upo.es}           
}

\date{Received: date / Accepted: date}

\maketitle

\begin{abstract}

A Poincar\'e-covariant continuum approach to the three valence-quark bound-state problem in quantum field theory is used to perform a detailed analysis of the nucleon's ground and first excited states: the so-called $N(940)\frac{1}{2}^+$ and $N(1440)\frac{1}{2}^+$. 
Such analysis predicts the presence of nonpointlike, fully-interacting quark-quark (diquark) correlations within them, being the isoscalar-scalar and isovector-pseudovector diquarks overwhelmingly dominant with similar relative strengths in both states. Moreover, the rest-frame wave functions of both states are largely $S$-wave in nature and the first excited state in this $1/2^+$ channel has the appearance of a radial excitation of the ground state.
All these features have numerous observable consequences, we show herein those related with the nucleon's elastic, Roper's elastic and nucleon-to-Roper transition electromagnetic form factors, for both charged and neutral channels.
%
%
\end{abstract}

\newpage

\section{Introduction}

The strong interaction is described by quantum Chromodynamics (QCD). This quantum field theory of gluons, as the gauge fields, and quarks, as the matter fields, is conceptually simple and can be expressed compactly in just one line with two definitions~\cite{Wilczek:2000ih}. However, nearly four decades after its formulation, we are still seeking answers to such apparently simple questions as what is the proton's wave function and which of the known baryons is the proton's first radial excitation. Indeed, numerous problems remain open because QCD is fundamentally different from other pieces of the Standard Model of Particle Physics: whilst perturbation theory is a powerful tool when used in connection with high-energy QCD processes, this technique is essentially useless when it comes to developing an understanding of strong interaction bound states built from light quarks.

Nonperturbative emergent phenomena such as confinement of gluons and quarks as well as dynamical chiral symmetry breaking (DCSB), appear to play a dominant role in determining all observable characteristics of QCD at low energy. The body of experimental and theoretical methods used to probe and map the infrared domain of QCD can be named as strong-QCD (sQCD). In this manuscript, we present a continuum formulation of the three valence-quark bound-state problem in quantum field theory in order to compute nucleon's ground and excited state properties: mass, wave function's peculiarities and associated elastic and transition electromagnetic form factors.

Whilst the proton is plainly a bound-state seeded by three valence-quarks: $uud$, and the neutron is similar; the $N(1440)\frac{1}{2}^+$ ``Roper resonance" has long been a source of puzzlement. This confusion was only resolved recently~\cite{Burkert:2017djo} with the acquisition and analysis of a vast amount of high-precision nucleon-resonance electro-production data with single- and double-pion final states~\cite{Aznauryan:2011qj} on a large kinematic domain of energy and photon virtuality, the development of a sophisticated dynamical reaction theory~\cite{Kamano:2016bgm, Ronchen:2015vfa} capable of simultaneously describing all partial waves extracted from available, reliable data, and the application of a Poincar\'e covariant approach to the continuum bound state problem in relativistic quantum field theory. This manuscript is intended to be a small revision of the last mentioned approach with particular emphasis on its derived results about elastic and transition electromagnetic form factors involving the nucleon and Roper.

A unified description of electromagnetic elastic and transition form factors involving the nucleon and its resonances has acquired very much interest. On the theoretical side, it is via the $Q^2$-evolution of form factors that one gains access to the running of QCD's coupling and masses~\cite{Cloet:2013gva, Chang:2013nia}. Moreover, QCD-based approaches that compute form factors at large photon virtualities are needed because the so-called meson-cloud often screens the dressed-quark core of all baryons at low momenta~\cite{Tiator:2003uu, Kamano:2013iva}. On the experimental side, a substantial progress has been made in the extraction of transition electrocouplings, $g_{{\rm v}NN^\ast}$, from meson electroproduction data, obtained primarily with the CLAS detector at the Jefferson Laboratory (JLab)~\cite{Aznauryan:2011qj, Tanabashi:2018oca,  Mokeev:2018zxt, Isupov:2017lnd, Mokeev:2013kka, Mokeev:2012vsa, Aznauryan:2012ba}. 

This manuscript is arranged as follows. We present in Sec.~\ref{sec:Faddeev} a short survey of our theoretical framework in order to compute the mass and wave function of the nucleon and its first excited state. Section~\ref{sec:EMcurrent} shows the way in which the associated baryon's electromagnetic current can be computed within our formalism. This section also contains our results on the elastic form factors of the nucleon and Roper resonance, both charged and neutral cases; and a dissection of the transition form factors associated with the $\gamma^\ast N \to R$ reaction. We finish in Sec.~\ref{sec:summary} giving some conclusions and an outlook of the work to be done in the following years.


\begin{figure}[!t]
\begin{center}
\hspace*{0.10cm}
\includegraphics[clip, width=0.90\textwidth, height=0.25\textwidth]{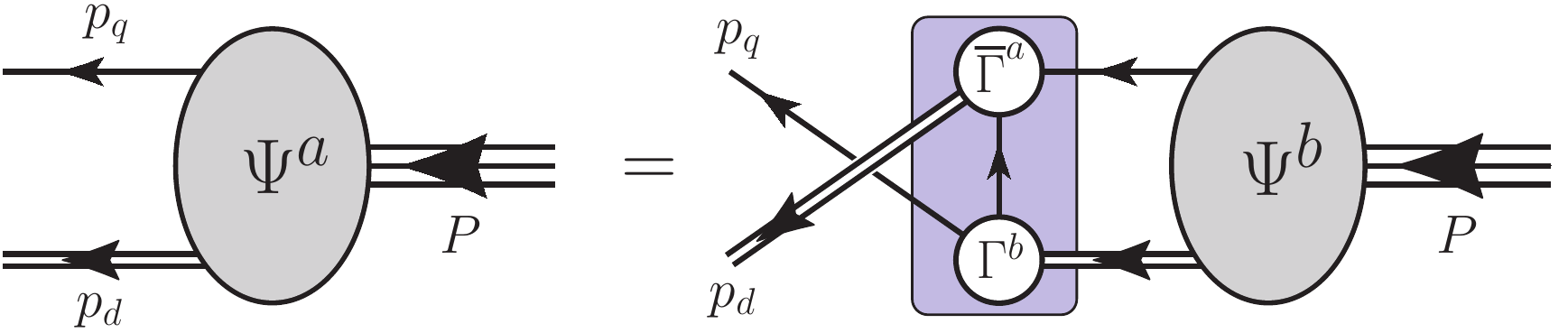}
\caption{\label{fig:Faddeev} Poincar\'e-covariant Faddeev equation: a homogeneous linear integral equation for the matrix-valued function $\Psi$, being the Faddeev amplitude for a baryon of total momentum $P = p_q + p_d$, which expresses the relative momentum correlation between the dressed-quarks and -diquarks within the baryon. The (purple) highlighted rectangle demarcates the kernel of the Faddeev equation: {\it single line}, dressed-quark propagator; {\it double line}, diquark propagator; and $\Gamma$, diquark correlation amplitude.
\vspace*{-0.50cm}
}
\end{center}
\end{figure}

\vspace*{-0.50cm}
\section{Nucleon's bound state problem}
\label{sec:Faddeev}

%
Two decades of studying three-body bound-state problems in hadron physics, e.g. Refs.~\cite{Oettel:1998bk, Roberts:2011cf, Eichmann:2016hgl, Eichmann:2016yit, Chen:2017pse, Lu:2017cln}, have evidenced the appearance of soft (nonpointlike) fully-interacting diquark correlations within baryons, whose characteristics are greatly influenced by DCSB~\cite{Segovia:2015ufa}. No realistic counter examples are known and the existence of such diquark correlations is also supported by lattice-regularised QCD simulations~\cite{Alexandrou:2006cq, Babich:2007ah}.

Consequently, the problem of determining the structure of the dressed-quark core of a baryon is transformed into that of solving the linear, homogeneous matrix equation depicted in Figure~\ref{fig:Faddeev} and introduced in Refs.~\cite{Cahill:1988dx, Burden:1988dt, Cahill:1988zi, Reinhardt:1989rw, Efimov:1990uz}, in combination with the realistic quark-quark interaction presented in Refs.~\cite{Binosi:2014aea, Binosi:2016wcx, Binosi:2016nme}. The Faddeev equation of Fig.~\ref{fig:Faddeev} involves three basic elements: (i) the dressed quark propagator, (ii) the propagator and correlation amplitude for all participating diquarks, and (iii) the Faddeev amplitude for a baryon. All these will be explained immediately below; herein, we only want to highlight that a baryon described by Fig.~\ref{fig:Faddeev} is a Borromean bound-state where the binding energy is given by two main contributions~\cite{Segovia:2015ufa}: One part is expressed in the formation of tight diquark correlations, the second one is generated by the quark exchange depicted in the highlighted rectangle of the Fig.~\ref{fig:Faddeev}. This  exchange ensures that no quark holds a special place because each one participates in all diquarks to the fullest extent allowed by its quantum numbers. The continual rearrangement of the quarks guarantees that the wave function complies with the baryon's fermionic nature.


\vspace*{-0.50cm}
\subsection{Dressed quark propagator}

An extensive literature~\cite{Lane:1974he, Politzer:1976tv, Zhang:2004gv, Bhagwat:2004kj, Bhagwat:2006tu, Binosi:2016wcx} exists about studying the dressed light-quark propagator:
\begin{equation}
S(p) = -i \gamma\cdot p\, \sigma_V(p^2) + \sigma_S(p^2) = \frac{1}{i\gamma\cdot p\, A(p^2) + B(p^2)} \,,
\end{equation}
showing that the wave function renormalisation, $Z(p^2)=1/A(p^2)$, and dressed-quark mass, $M(p^2)=B(p^2)/A(p^2)$, receive strong momentum-dependent corrections at infrared momenta: $Z(p^2)$ is suppressed whereas $M(p^2)$ is enhanced.

An efficacious parametrisation of $S(p)$, which exhibits the features described above, has been used extensively in hadron studies~\cite{Roberts:2007jh}. It is expressed via
\begin{align}
\bar\sigma_S(x) &= 2\,\bar m \,{\cal F}(2 (x+\bar m^2)) + {\cal F}(b_1 x) \,{\cal F}(b_3 x) \, \left[b_0 + b_2 {\cal F}(\epsilon x)\right] \,, \label{ssm} \\
\bar\sigma_V(x) & =  \frac{1}{x+\bar m^2}\, \left[ 1 - {\cal F}(2
(x+\bar m^2))\right] \label{svm} \,,
\end{align}
with $x=p^2/\lambda^2$, $\bar m$ = $m/\lambda$,
\begin{equation}
\label{defcalF}
{\cal F}(x)= \frac{1-\mbox{\rm e}^{-x}}{x}  \,,
\end{equation}
$\bar\sigma_S(x) = \lambda\,\sigma_S(p^2)$ and $\bar\sigma_V(x) =
\lambda^2\,\sigma_V(p^2)$. The mass-scale, $\lambda=0.566\,$GeV, and parameter values
\begin{equation}
\label{tableA}
\begin{array}{ccccccccc}
\bar m  & \; & b_0   & \; & b_1  & \; & b_2   & \; & b_3 \\
\hline
0.00897 & \; & 0.131 & \; & 2.90 & \; & 0.603 & \; & 0.185
\end{array}
\end{equation}
The dimensionless $u=d$ current-quark mass in Eq.\,(\ref{tableA}) corresponds to
$m=5.08\,{\rm MeV}$ and the parametrisation yields the following Euclidean constituent-quark mass, defined as the solution of $p^2=M^2(p^2)$: $M_{u,d}^E = 0.33\,{\rm GeV}$. The ratio $M^E/m = 65$ is an expression of DCSB in the parametrisation of $S(p)$. It emphasises the dramatic enhancement of the dressed-quark mass function at infrared momenta. It is also important to note that Eqs.~\eqref{ssm} and~\eqref{svm} ensure confinement of the dressed quarks via the violation of reflection positivity~\cite{Horn:2016rip}.

\vspace*{-0.50cm}
\subsection{Diquark correlation}

Five types of diquark correlations are possible within a baryon with quantum numbers $(I,J^P)=(1/2,1/2^+)$: isoscalar-scalar, isovector-pseudo\-vector, isoscalar-pseudo\-scalar, iso\-scalar-vector, and iso\-vector-vector. However, re\-fe\-ren\-ces~\cite{Chen:2012qr, Eichmann:2016hgl, Lu:2017cln, Chen:2017pse} have de\-mons\-trated that the dominant diquark correlations for the nucleon's ground and first excited states are isoscalar-scalar $(0,0^+)$ and isovector-pseudovector $(1,1^+)$ diquarks. The leading structure of the diquark correlation amplitude is given, for each case, by:
\begin{align}
\label{eq:DiqAmp1}
\Gamma^{0^+}(k;K) & = g_{0^+} \, \gamma_5 C\, \tau^2 \,\bm{H} \,{\cal F}(k^2/\omega_{0^+}^2) \,, \\
\label{eq:DiqAmp2}
\bm{\Gamma}_\mu^{1^+}(k;K) & = i g_{1^+} \, \gamma_\mu C \, \bm{t}\, \bm{H} \,{\cal F}(k^2/\omega_{1^+}^2) \,,
\end{align}
where $k$ is a two-body relative momentum, $K$ is the total momentum of the correlation,  ${\cal F}$ is the function in Eq.~\eqref{defcalF}, $g_{J^P}$ is a coupling into the diquark channel which is fixed by canonical normalization and modulated by the size parameter $\omega_{J^P}$:
\begin{equation}
\omega_{J^P} = \frac{m_{J^P}}{\sqrt{2}} \,.
\end{equation}
The mass-scales, which express the strength and range of the correlation, have been constrained by numerous studies~\cite{Chen:2012qr, Lu:2017cln, Chen:2017pse}; we use
\begin{equation}
m_{0^+} = 0.8\,\text{GeV} \,, \quad m_{1^+} = 0.9\,\text{GeV} \,,
\end{equation}
and we find
\begin{equation}
g_{0^+} = 14.8 \,, \quad g_{1^+} = 12.7 \,.
\end{equation}
About the matrices which appear in Eqs.~\eqref{eq:DiqAmp1} and~\eqref{eq:DiqAmp2}, $C=\gamma_2\gamma_4$ is the charge-conjugation matrix; $\{t^j, j=+,0,-\} = \tfrac{1}{\surd 2} \{(\tau^0+\tau^3), \surd 2 \, \tau^1, (\tau^0-\tau^3)\}$, $\tau^0=\,$diag$[1,1]$, $\{\tau^i,i=1,2,3\}$ are the Pauli matrices; and $H = \{i\lambda_c^7, -i\lambda_c^5,i\lambda_c^2\}$, with $\{\lambda_c^k,k=1,\ldots,8\}$ denoting Gell-Mann matrices in colour space, expresses the diquarks' colour antitriplet character. Note herein that the colour-sextet quark+quark channel does not support correlations because gluon exchange is repulsive.

We associate a propagator to each quark-quark correlation in Fig.~\ref{fig:Faddeev}. For the scalar and pseudovector diquarks, the propagator is given by the following expressions~\cite{Segovia:2014aza}:
\begin{subequations}
\label{Eqqqprop}
\begin{align}
\Delta^{0^+}(K) & = \frac{1}{m_{0^+}^2} \, {\cal F}(k^2/\omega_{0^+}^2)\,,\\
\Delta^{1^+}_{\mu\nu}(K) & = \left[ \delta_{\mu\nu} + \frac{K_\mu K_\nu}{m_{1^+}^2} \right] \frac{1}{m_{1^+}^2} \, {\cal F}(k^2/\omega_{1^+}^2)\,.
\end{align}
\end{subequations}
These algebraic forms ensure that the diquarks are confined within the baryons, as appropriate for coloured correlations: whilst the propagators are free-particle-like at spacelike momenta, they are pole-free on the timelike axis; and this is sufficient to ensure confinement via the violation of reflection positivity~\cite{Horn:2016rip}.


\vspace*{-0.50cm}
\subsection{Faddeev amplitude}

The nucleon, and its first excited state, can be represented by the Faddeev amplitude
\begin{equation}
\Psi = \psi_{1} + \psi_{2} + \psi_{3}  \,,
\label{PsiNucleon}
\end{equation}
where the subscript identifies the bystander quark, {\it i.e.} the quark that is not participating in a diquark correlation. Denoting $\psi_{3} \equiv \psi$, we have:
\begin{align}
\nonumber
\psi (p_i,\sigma_i,\alpha_i) &= [\Gamma^{0^+}(k;K)]^{\alpha_1 \alpha_2}_{\sigma_1 \sigma_2} \, \Delta^{0^+}(K) \,[\varphi_{0^+}(\ell;P) u(P)]^{\alpha_3}_{\sigma_3} \nonumber \\
&
+ [\Gamma^{1^+j}_\mu] \, \Delta_{\mu\nu}^{1^+}\, [\varphi_{1^+ \nu }^{j}(\ell;P) u(P)] \,,
\label{FaddeevAmp}
\end{align}
where $(p_i,\sigma_i,\alpha_i)$ are the momentum, spin and isospin labels of the quarks constituting the bound state; $P=p_1 + p_2 + p_3 = p_q + p_d$ is the total momentum of the baryon; $k=(p_1-p_2)/2$, $K=p_1+p_2=p_d$, $\ell=(-K + 2 p_3)/3$; $j$ is the flavour label in the Pauli matrices defined in the former subsection; and $u(P)$ is a Euclidean spinor (see Ref.\,\cite{Segovia:2014aza}, Appendix~B for details). The matrix-valued functions in Eq.~\eqref{FaddeevAmp} that are not yet defined are given by:
\begin{align}
\varphi_{0^+}(\ell;P) & = \sum_{i=1}^2 {\mathpzc s}_i(\ell^2,\ell\cdot P)\,  {\cal S}^i(\ell;P) \,, \label{eq:scalars1} \\
\varphi_{1^+ \nu}^{j}(\ell;P)  & = \sum_{i=1}^6 {\mathpzc a}_i^{j}(\ell^2,\ell\cdot P)\, \gamma_5 {\cal A}^i_\nu(\ell;P) \,, \label{eq:scalars2}
\end{align}
where
\begin{center}
\begin{tabular}{lll}
$\,{\cal S}^1 = {\mathbf I}_{\rm D} \,,$ &
$\,{\cal S}^2 = i \gamma\cdot\hat\ell - \hat\ell \cdot\hat P {\mathbf I}_{\rm D} \,,$ & \\
${\cal A}^1_\nu =  \gamma\cdot\ell^\perp \hat P_\nu \,,$ &
${\cal A}^2_\nu = - i \hat P_\nu {\mathbf I}_{\rm D} \,,$ &
${\cal A}^3_\nu = \gamma\cdot\hat\ell^\perp \hat\ell^\perp_\nu \,,$ \\
${\cal A}^4_\nu = i\hat \ell_\nu^\perp {\mathbf I}_{\rm D} \,,$ &
${\cal A}^5_\nu = \gamma_\nu^\perp - {\cal A}^3_\nu \,,$ &
${\cal A}^6_\nu = i \gamma_\nu^\perp \gamma\cdot\hat\ell^\perp - {\cal A}^4_\nu \,,$
\end{tabular}
\end{center}
with $\hat\ell^2=1$, $\hat P^2 = -1$, $\ell_\nu^\perp = \hat\ell_\nu +\hat\ell\cdot\hat P \hat P_\nu$, $\gamma_\nu^\perp = \gamma_\nu +\gamma\cdot\hat P \hat P_\nu$.


\vspace*{-0.50cm}
\subsection{Solution of the nucleon's ground and first excited states}

Our computed values for the mass of the nucleon and its first excited state are~\cite{Chen:2017pse}:
\begin{equation}
\label{eqMasses}
\mbox{nucleon\,(N)} = 1.19\,\text{GeV} \,, \quad\quad \mbox{nucleon-excited\,(R)} = 1.73\,\text{GeV} \,.
\end{equation}
The empirical values of the pole locations for the first two states in the nucleon channel are~\cite{Tanabashi:2018oca, Suzuki:2009nj}: $0.939\,\text{GeV}$ and $(1.36 - i \, 0.091)\,\text{GeV}$, respectively. At first glance, these values appear unrelated to those shown in Eq.~\eqref{eqMasses}. However, deeper consideration~\cite{Eichmann:2008ae, Eichmann:2008ef} reveals that the kernel in Fig.~\ref{fig:Faddeev} omits all those resonant contributions which may be associated with the meson-baryon final-state interactions (MB\,FSIs) that are resummed in dynamical coupled channels models~\cite{Suzuki:2009nj, Kamano:2013iva, Doring:2014qaa} in order to transform a bare-baryon into the observed state. Our Faddeev equation should therefore be understood as producing the dressed-quark core of the bound-state, not the completely-dressed and hence observable object. In consequence, a comparison between the empirical values of the resonance pole positions and the masses in Eq.~\eqref{eqMasses} is not pertinent. Instead, one should compare the masses of the quark core with values determined for the meson-undressed bare-excitations, e.g.: 
\begin{center}
\begin{tabular}{l|ccccccc}
& herein & $\quad$ & \cite{Lu:2017cln} & $\quad$ & \cite{Suzuki:2009nj} & $\quad$ & \cite{Gasparyan:2003fp} \\
\hline
$M_N\,\text{(GeV)}$ & $1.19$ & & $1.14$ & & $-$    & & $1.24$ \\
$M_R\,\text{(GeV)}$ & $1.73$ & & $1.82$ & & $1.76$ & & $-$
\end{tabular}
\end{center}
The bare Roper mass in Ref.~\cite{Suzuki:2009nj} agrees with both our quark-core result and that obtained using a refined treatment of a vector$\,\otimes\,$vector contact-interaction~\cite{Lu:2017cln}. This is notable because all these calculations are independent, with just one common feature; namely, an appreciation that measured hadrons can realistically be built from a dressed-quark core plus a meson-cloud.

The Faddeev amplitude and wave function are Poincar\'e covariant. This means that the scalar functions that appear in Eqs.~\eqref{eq:scalars1} and~\eqref{eq:scalars2} are frame-independent. However, the frame chosen determines how these functions should be combined in order to guess the $L$- and $S$-nature of a particular quark-diquark channel in a given baryon. Details can be extracted from Table~I and Fig.~2 of Ref.~\cite{Chen:2017pse}, herein we only want to collect the bound-state mass depending on the chosen partial waves that compose the baryon in its rest-frame:
\begin{equation}
\label{Pwave}
\begin{array}{c|cccc}
\text{L content} \hspace*{0.20cm} & \quad & M_N\,\text{(GeV)} & \quad & M_R\,\text{(GeV)} \\
\hline
S, P, D   & & 1.19 & & 1.73 \\
\hline
-, P, D   & & -    & & -    \\
S, -, D   & & 1.24 & & 1.71 \\
S, P, -   & & 1.20 & & 1.74 \\
\end{array}
\end{equation}
Plainly, the nucleon and $N(1440)\frac{1}{2}^+$ are primarily $S$-wave in nature, since they are not supported by the Faddeev equation unless $S$-wave components are contained in the wave function.

\begin{figure}[!t]
\centerline{%
\includegraphics[clip,width=0.45\linewidth,height=0.25\textheight]{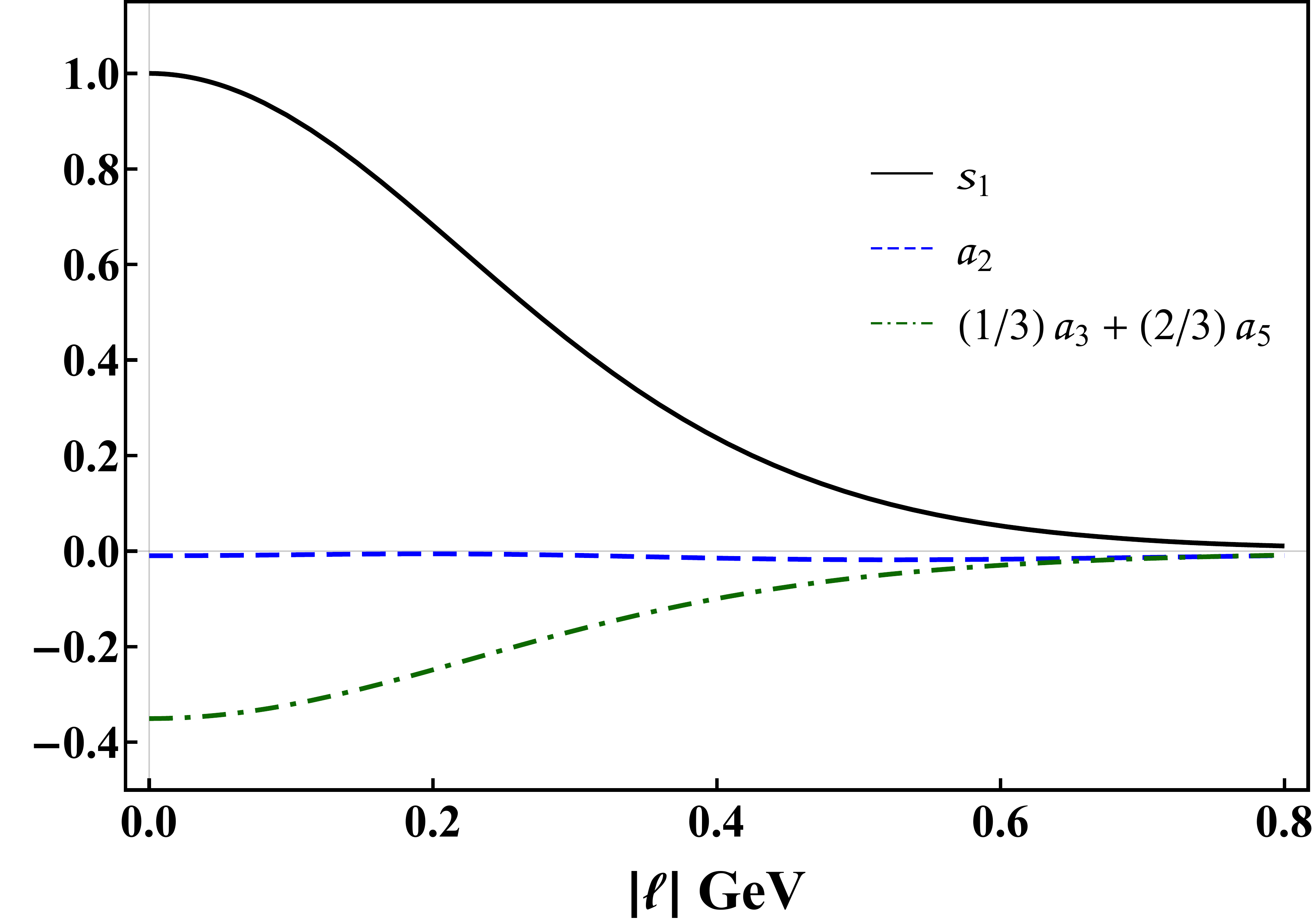}
\hspace*{0.40cm}
\includegraphics[clip,width=0.45\linewidth,height=0.25\textheight]{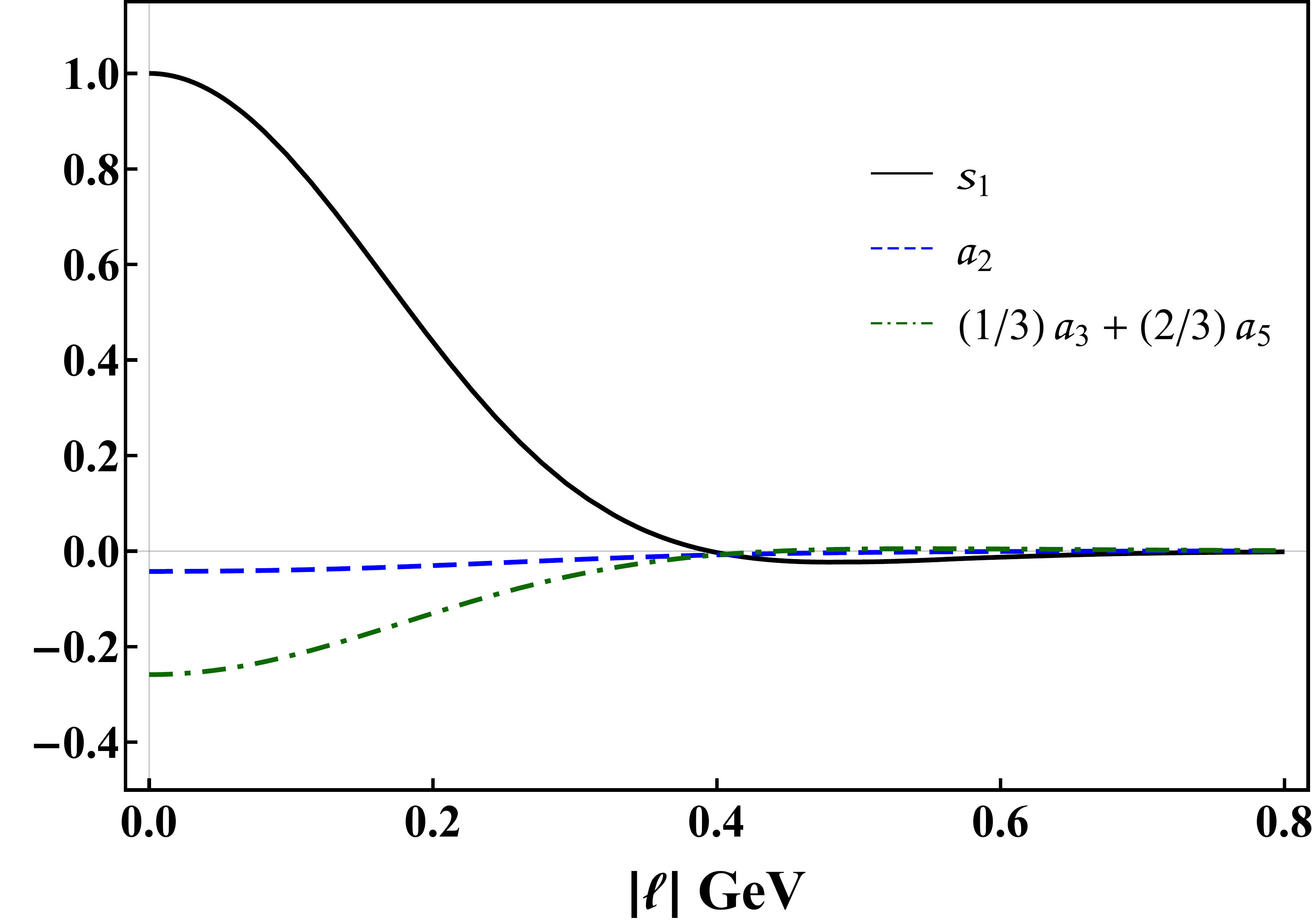}
}
\caption{\label{figFA}
\emph{Left panel}. Zeroth Chebyshev moment of all $S$-wave components in the nucleon's Faddeev wave function, which is obtained from $\Psi$ in Fig.~\ref{fig:Faddeev}, by reattaching the dressed-quark and -diquark legs.
\emph{Right panel}. Kindred functions for the first excited state.
Legend: ${\mathpzc s}_1$ is associated with the baryon's scalar diquark; the other two curves are associated with the axial-vector diquark; and the normalisation is chosen such that ${\mathpzc s}_1(0)=1$.  Details are provided in Refs.~\cite{Chen:2017pse, Segovia:2014aza}.
\vspace*{-0.50cm}
}
\end{figure}

Figure~\ref{figFA} shows now the order-zero Chebyshev projections of all $S$-wave components in the Faddeev wave function of the nucleon and its positive-parity excitation:
\begin{equation}
{\mathpzc W}(\ell^2;P^2) = \frac{2}{\pi} \int_{-1}^1 \! dx\,\sqrt{1-x^2}\,
{\mathpzc W}(\ell^2,x; P^2)\,,
\end{equation}
where $x=\ell\cdot P/\sqrt{\ell^2 P^2}$. Evidently, whilst these projections of the nucleon's Faddeev wave function are each of a single sign, either positive or negative, those associated with the quark core of the nucleon's first positive-parity excitation are quite different: all $S$-wave components exhibit a single zero at $z_R \approx 0.4\,$GeV$\approx 1/[0.5\,{\rm fm}]$. Drawing upon experience with quantum mechanics and with excited-state mesons studied via the Bethe-Salpeter equation~\cite{Holl:2004fr, Qin:2011xq, Rojas:2014aka, Li:2016dzv, Li:2016mah}, this pattern of behavior for the first excited state indicates that it may be interpreted as a radial excitation.

Based on the Faddeev amplitude's canonical normalisation, which is a Poincar\'e invariant quantity related to baryon number, the diquark fractions of the nucleon and its radial excitation can be summarize as follows:
\begin{equation}
\label{Pdiquark}
\begin{array}{llcccc}
        & & N    & & R    \\
\hline
P_{J=0} & $\quad$ & 62\% & $\quad$ & 62\% \\
P_{J=1} &         & 38\% &         & 38\% \\
\end{array}\,;
\end{equation}
namely, the relative strength of isoscalar-scalar and isovector-pseudovector diquark correlations in the nucleon and its radial excitation is the same. This conclusion was also found in Ref.~\cite{Chen:2017pse} but using a different scheme. Taking into account the values reported in Eq.~\eqref{Pdiquark}, Mezrag {\it et al.}~\cite{Mezrag:2017znp} reported recently the first quantum field theory calculation of the pointwise behaviour of the leading-twist parton distribution amplitudes (PDAs) of the proton and Roper.


\begin{figure}[!t]
\centerline{\includegraphics[clip,width=0.95\linewidth]{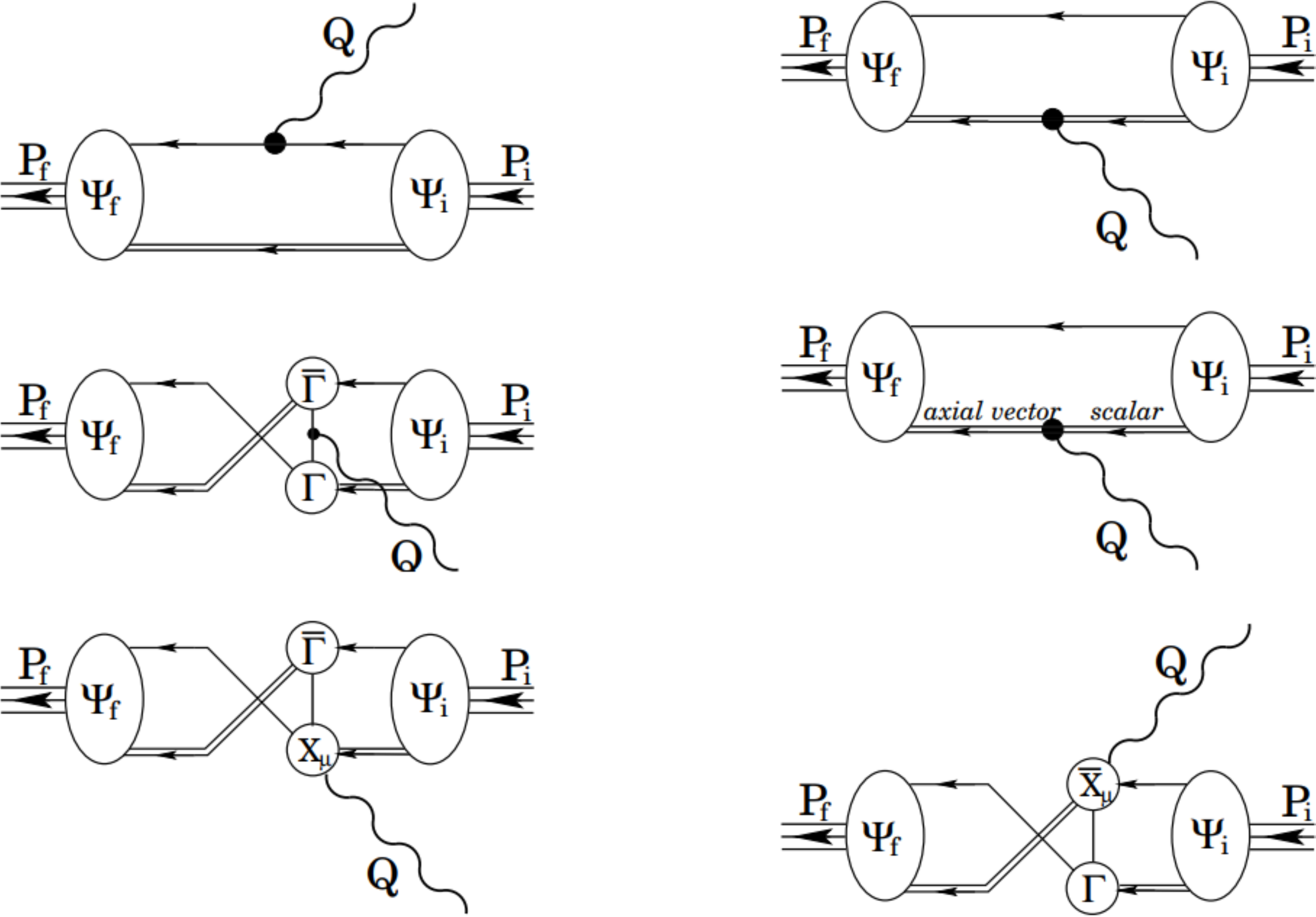}}
\caption{\label{vertexB} Vertex that ensures a conserved electromagnetic current for on-shell baryons that are described by the Faddeev amplitudes produced by the equation depicted in Fig.~\ref{fig:Faddeev}: \emph{single line}, dressed-quark propagator; \emph{undulating line}, photon; $\Gamma$,  diquark correlation amplitude; and \emph{double line}, diquark propagator. 
Diagram~1 is the top-left image; the top-right is Diagram~2; and so on, with Diagram~6 being the bottom-right image.
Details related with the explicit calculation of each diagram can be found in Ref.~\cite{Segovia:2014aza}, Appendix~C.
\vspace*{-0.50cm}
}
\end{figure}

\vspace*{-0.50cm}
\section{Nucleon's electromagnetic form factors}
\label{sec:EMcurrent}

Within our formalism, the calculation of the desired elastic and transition electromagnetic form factors is straightforward once the corresponding current is specified. The electromagnetic current between initial and final states with quantum numbers $(I,J^P)=(1/2,1/2^+)$ is completely specified by two form factors:
\begin{equation}
\bar u_{f}(P_f) \Big[ \gamma_\mu^T F_{1}^{fi}(Q^2)+\frac{1}{m_{{fi}}} \sigma_{\mu\nu} Q_\nu F_{2}^{fi}(Q^2) \Big] u_{i}(P_i)\,,
\label{NRcurrents}
\end{equation}
where $u_{i}$, $\bar u_{f}$ are, respectively, Dirac spinors describing the incoming/outgoing baryons, with four-momenta $P_{i,f}$ and masses $m_{i,f}$ so that $P_{i,f}^2=-m_{i,f}^2$; $Q=P_f-P_i$; $m_{{fi}} = (m_f+m_{i})$; and $\gamma^T \cdot Q= 0$.

References~\cite{Oettel:1999gc, Segovia:2014aza} contain detailed information on the interaction of a photon with a baryon generated by the Faddeev equation depicted in Fig.~\ref{fig:Faddeev}. There are six terms, shown in Fig.~\ref{vertexB}, in which the photon separately probes the quarks and diquarks in various ways, so that diverse features of quark dressing and quark-quark correlations play a role in determining the form factors.


\vspace*{-0.50cm}
\subsection{Elastic form factors}

\begin{figure}[!t]
\centerline{%
\hspace*{-0.10cm}
\includegraphics[clip, width=0.45\textwidth]{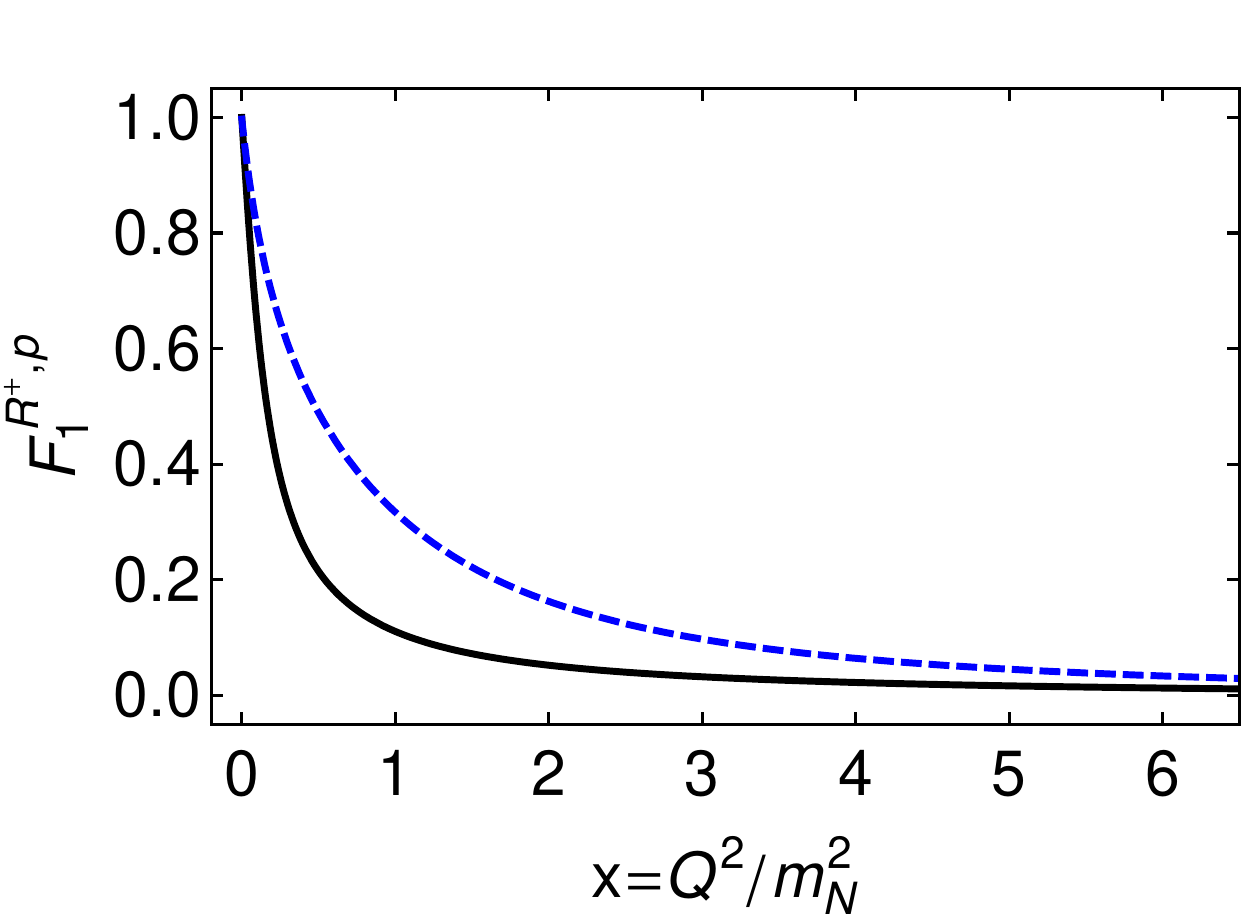} \hspace*{0.10cm}
\includegraphics[clip, width=0.475\textwidth, height=0.205\textheight]{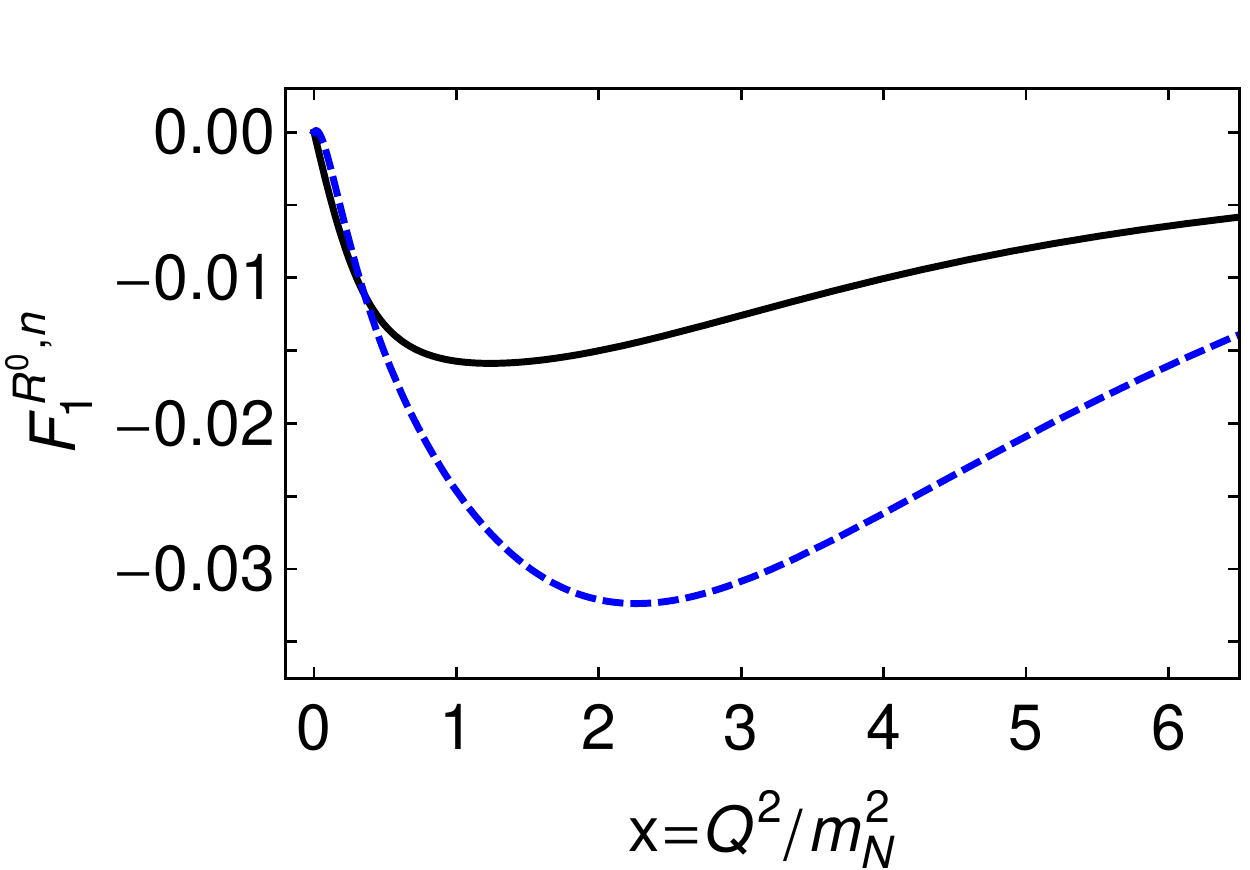} \\
}
\centerline{%
\includegraphics[clip, width=0.45\textwidth]{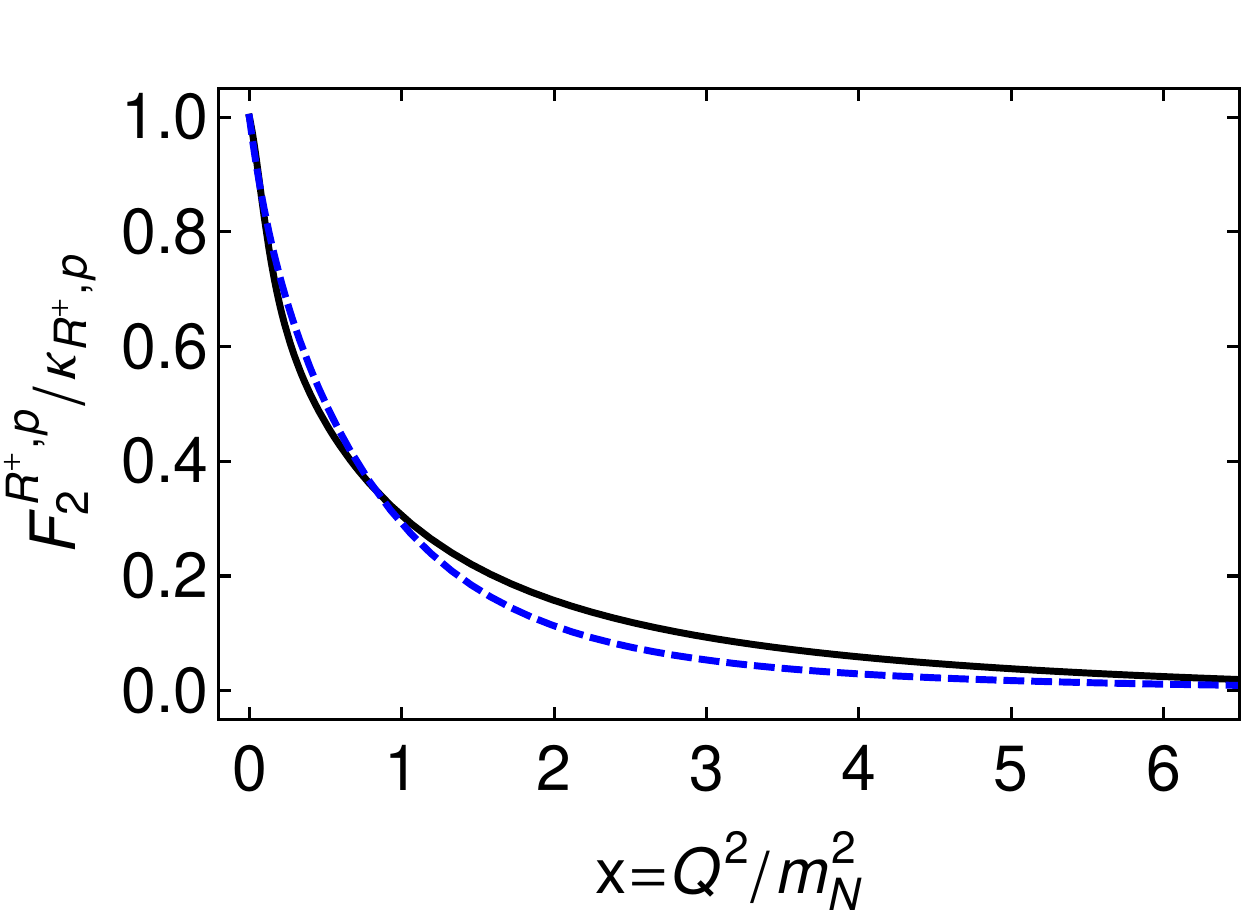} \hspace*{0.40cm}
\includegraphics[clip, width=0.45\textwidth]{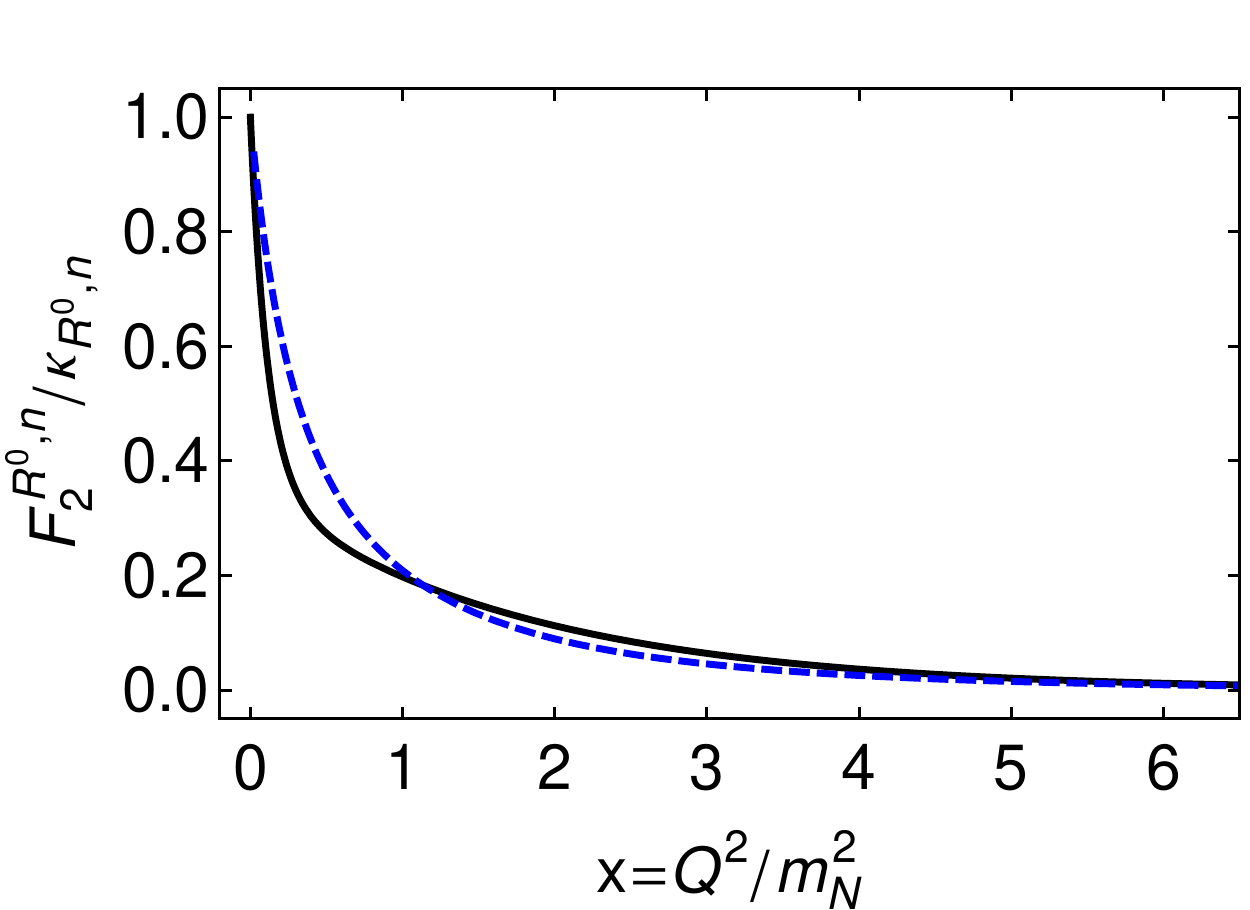}
}
\caption{\label{elastic}
Solid (black) curves -- Dirac (upper panels) and Pauli (lower panels) elastic electromagnetic form factors associated with the dressed-quark cores of the charged (left panels) and neutral (right panels) Roper systems.
Dashed (blue) curves -- analogous results for the nucleon's ground state.
$\kappa_{R,N} = F_2^{R,N}(x=0)$ and $x=Q^2/m_N^2$, where $M_N = 1.18\,\text{GeV}$ is the dressed-quark core mass of the nucleon.
\vspace*{-0.50cm}
}
\end{figure}

Figure~\ref{elastic} shows the Dirac and Pauli elastic electromagnetic form factors of the nucleon and Roper. The panels on the left refer to the charged case whereas the ones on the right show the neutral one. Evidently, there are qualitative similarities and quantitative differences between the results of the nucleon and Roper. The biggest difference appear in the $Q^2$-dependence of the Dirac form factors, particularly striking appear to be the dissimilarity between neutral-Roper and neutron; but here appearances are deceptive because both functions are independently computed as the valence-quark electric-charge-weighted sum of larger, positive quantities, with cancellations leading to small results.

The (Sachs) electric and magnetic form factors can be defined as
\begin{align}
G_E(Q^2) = F_1(Q^2) - \frac{Q^2}{4 m_B^2} F_2(Q^2)\,,\quad
G_M(Q^2) = F_1(Q^2) + F_2(Q^2)\,,
\end{align}
where $m_B$ is the baryon's mass. The $Q^2=0$ values and slopes of the Sachs electric and magnetic form factors yield the static properties listed in Table~\ref{tabstatic}, where the radii are defined as
\begin{align}
\label{eqradii}
r^2 & = - \left.\frac{6}{\mathpzc n} \frac{d}{ dQ^2} G(Q^2)\right|_{Q^2=0}\,,
\end{align}
with ${\mathpzc n} = G(Q^2=0)$ when this quantity is nonzero, ${\mathpzc n} = 1$ otherwise, and the anomalous magnetic moment $\mu = G_M(0)$. The electromagnetic radii of the charged-Roper core are larger than those of the proton core, but the magnetic moments are similar; and this pattern is reversed in the neutral-Roper/neutron comparison.

\begin{table}[!t]
\caption{\label{tabstatic} Static properties {\it derived} from the elastic form factors depicted in Fig.~\ref{elastic}, see Eq.~\eqref{eqradii} and following text. The nucleon dressed-quark core mass is $M_N = 1.18\,\text{GeV}$.}
\begin{center}
\begin{tabular*}
{\hsize}
{
l|@{\extracolsep{0ptplus1fil}}
c@{\extracolsep{0ptplus1fil}}
c@{\extracolsep{0ptplus1fil}}
c@{\extracolsep{0ptplus1fil}}
c@{\extracolsep{0ptplus1fil}}}\hline
  & $R^+$ & $p$ & $R^0$ & $n$ \\\hline
$r_E \, M_N$ & 6.23 & 3.65 & $\phantom{-}0.93i$ & $\phantom{-}1.67i$\\
$r_M \, M_N$ & 4.49 & 3.17 & $\phantom{-}4.15\phantom{i}$ & $\phantom{-}4.19\phantom{i}$\\
$\mu$ & 2.67 & 2.50 & $-1.24\phantom{i}$ & $-1.83\phantom{i}$ \\\hline
\end{tabular*}
\end{center}
\vspace*{-0.50cm}
\end{table}


\vspace*{-0.50cm}
\subsection{Transition form factors}

Transition electromagnetic form factors involving the nucleon and Roper may be dissected in two separate ways, each of which can be considered as a sum of three distinct terms, \emph{viz}.
\begin{itemize}
\item Diquark dissection:
\vspace*{0.20cm}
\begin{itemize}
\item[] DD1.- Scalar diquark in both the initial and final baryon.
\vspace*{0.10cm}
\item[] DD2.- Pseudovector diquark in both the initial and final baryon.
\vspace*{0.10cm}
\item[] DD3.- A different diquark in the initial and final baryon.
\end{itemize}
\vspace*{0.20cm}
\item Scatterer dissection:
\vspace*{0.20cm}
\begin{itemize}
\item[] DS1.- Photon strikes a bystander quark.
\vspace*{0.10cm}
\item[] DS2.- Photon interacts with a diquark, elastically or causing a transition between scalar and pseudovector cases.
\vspace*{0.10cm}
\item[] DS3.- Photon strikes a dressed-quark in-flight, as one diquark breaks up and another is formed, or appears in one of the two associated ``seagull'' terms.
\end{itemize}
\end{itemize}

\begin{figure}[!t]
\begin{center}
\begin{tabular}{cc}
\includegraphics[clip,width=0.45\linewidth]{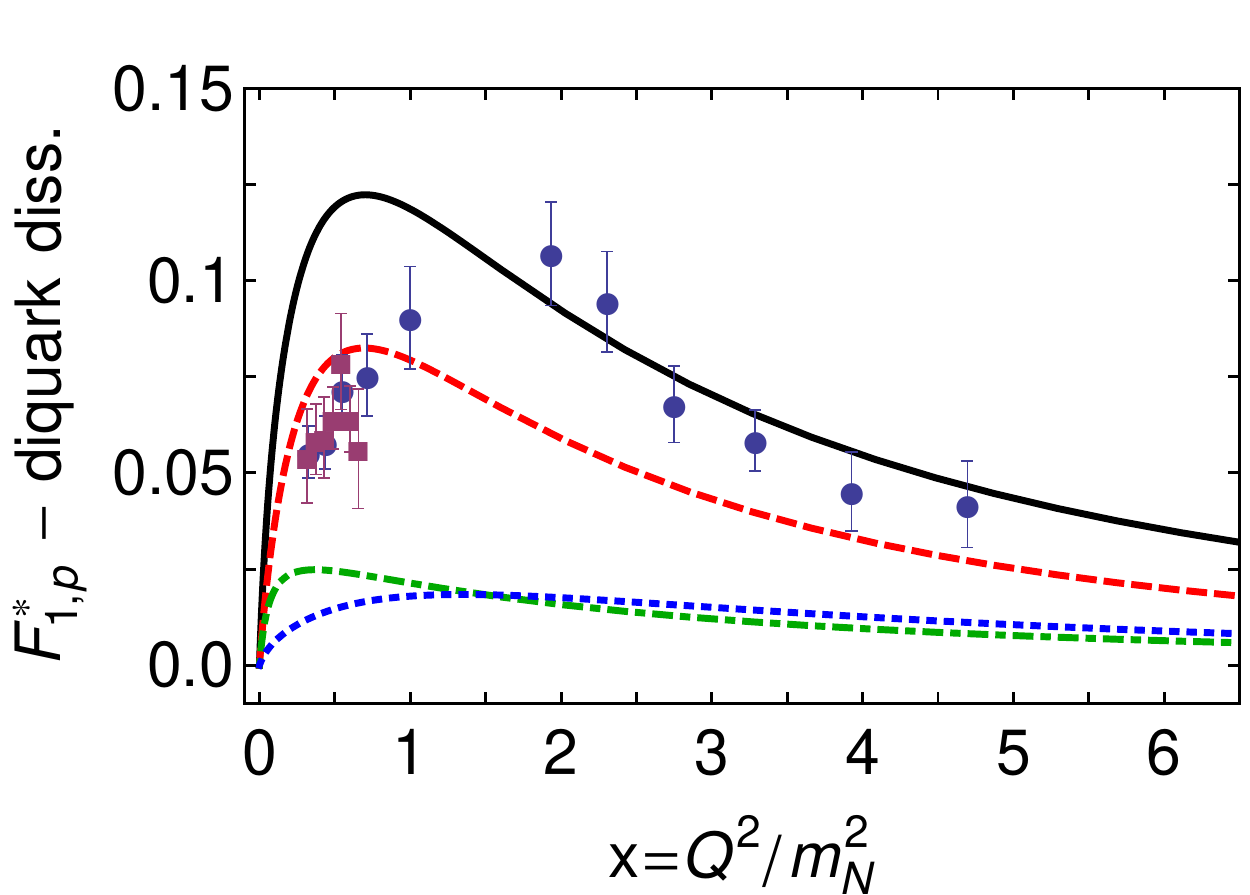} &
\includegraphics[clip,width=0.45\linewidth]{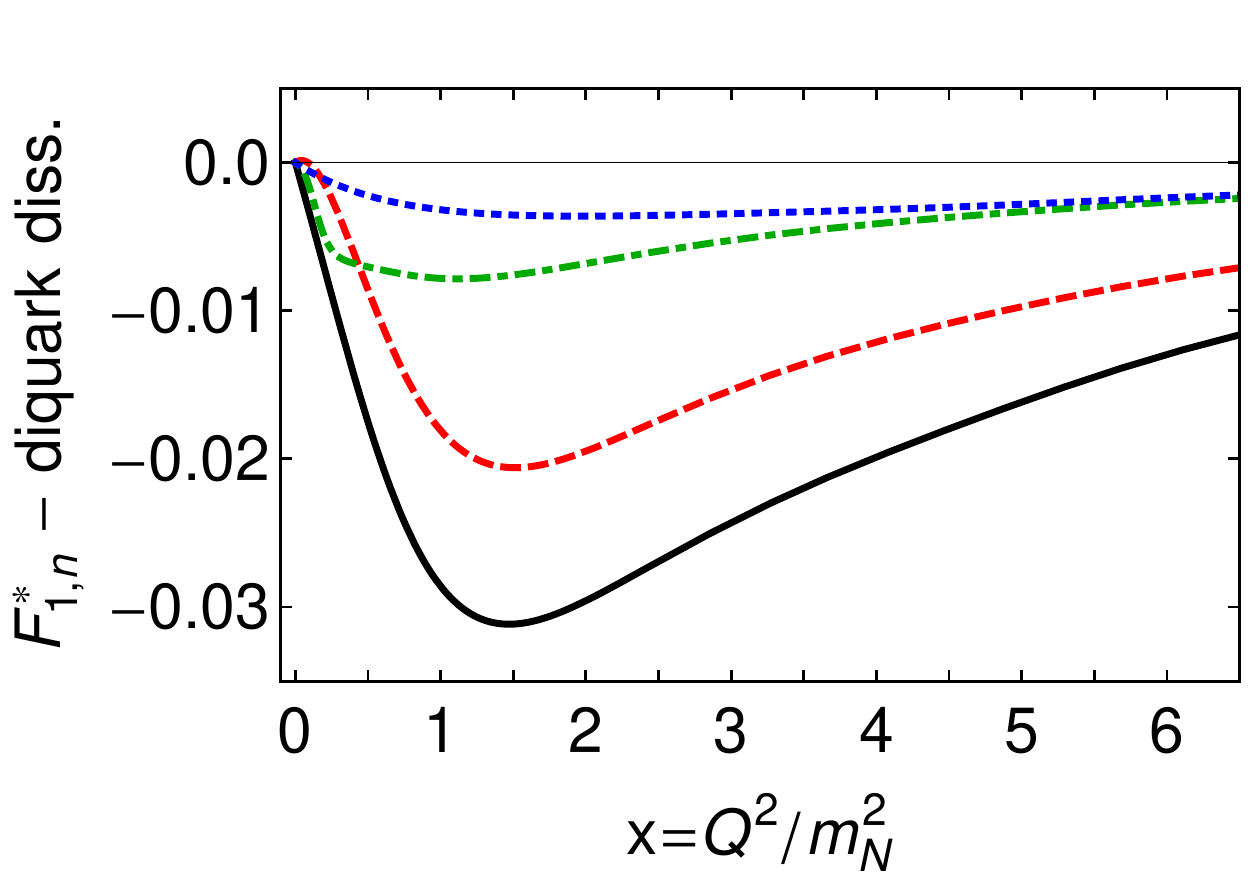}
\end{tabular}
\begin{tabular}{cc}
\includegraphics[clip,width=0.45\linewidth]{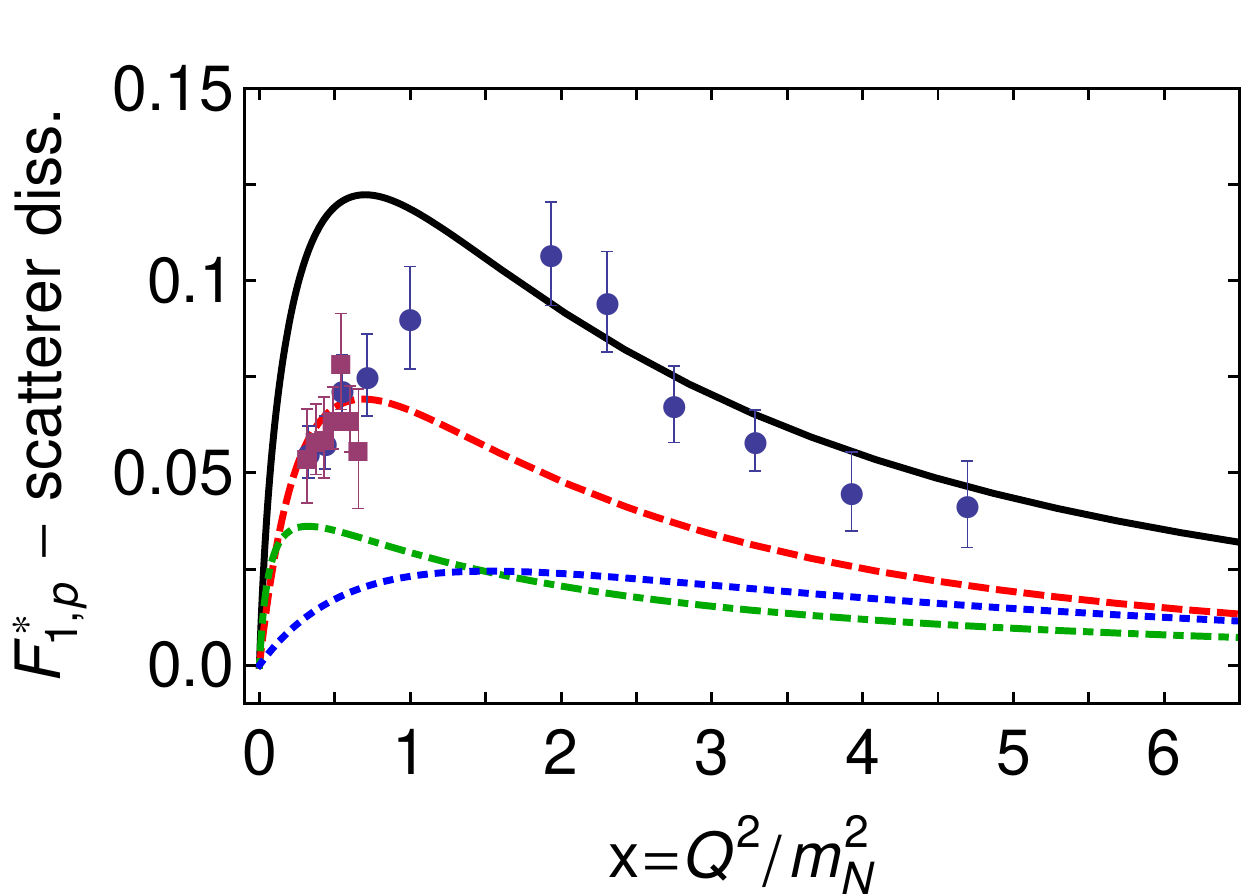} &
\includegraphics[clip,width=0.45\linewidth]{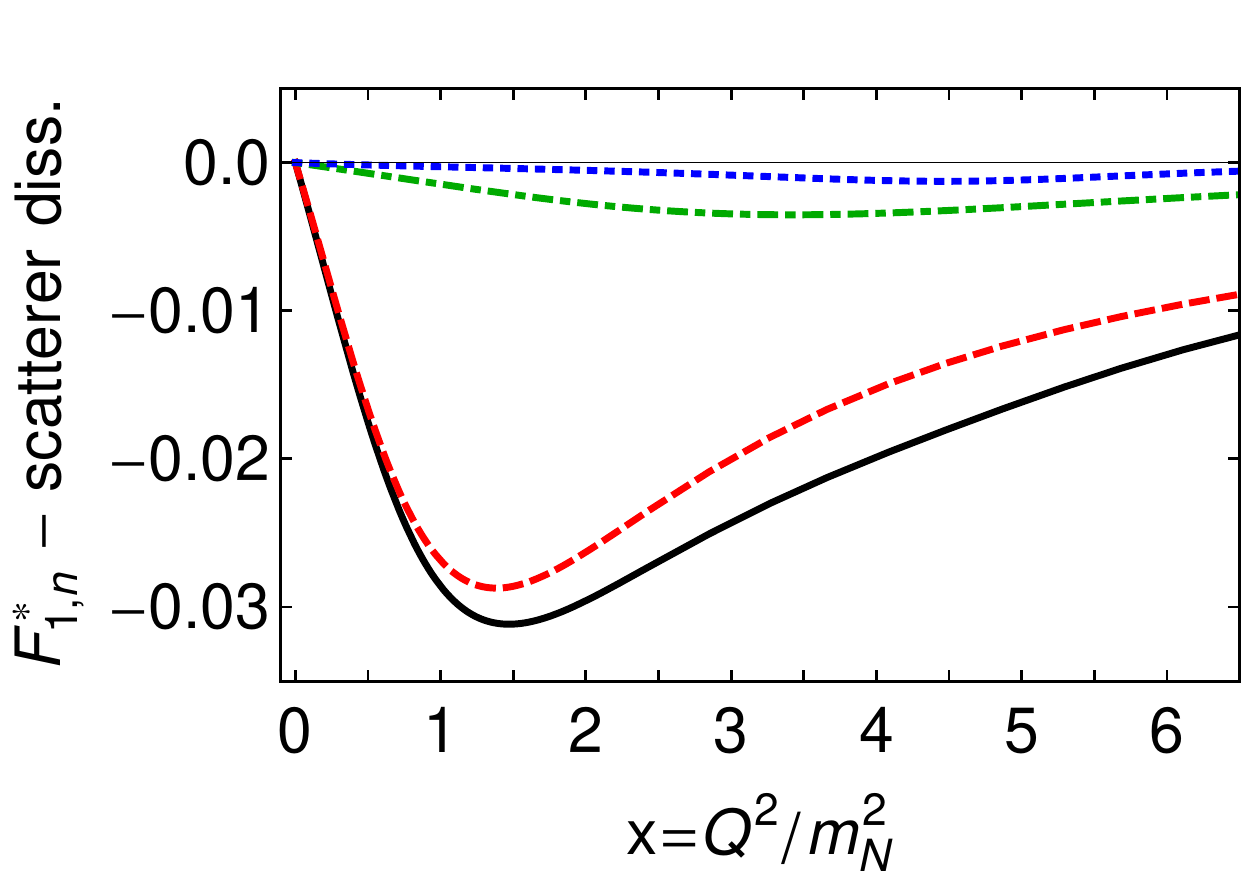}
\end{tabular}
\end{center}
\caption{\label{figF1}
Computed Dirac transition form factor, $F_{1}^{\ast}$, for the charged reaction $\gamma^\ast\,p\to R^+$ (left panels) and the neutral reaction $\gamma^\ast\,n\to R^0$ (right panels): solid (black) curve in each panel.
Data, left panels: circles (blue)~\cite{Aznauryan:2009mx}, and squares (purple)~\cite{Mokeev:2012vsa, Mokeev:2015lda}.
\emph{Upper panels} -- diquark breakdown: \emph{DD1} (dashed red), scalar diquark in both nucleon and Roper; \emph{DD2} (dot-dashed green), pseudovector diquark in both nucleon and Roper; \emph{DD3} (dotted blue), scalar diquark in nucleon, pseudovector diquark in Roper, and vice versa.
\emph{Lower panels} -- scatterer breakdown: \emph{DS1} (red dashed), photon strikes an uncorrelated dressed-quark; \emph{DS2} (dot-dashed green), photon strikes a diquark; and \emph{DS3} (dotted blue), diquark breakup contributions, including photon striking exchanged dressed-quark.
\vspace*{-0.50cm}
}
\end{figure}

Our predictions for the $\gamma^\ast N \to R$ Dirac transition form factors are drawn in Fig.~\ref{figF1}. They must vanish at $x=0$ owing to orthogonality between the nucleon and its radial excitation. Plainly, the charged transition proceeds primarily through a photon striking a bystander dressed-quark that is partnered by $[ud]$, with lesser but non-negligible contributions from all other processes. The neutral transition proceeds also primarily through a photon striking a bystander dressed-quark that is partnered by $[ud]$. Herein, it is important to highlight that charge neutrality enforces $F_1^{R^0}(0)=0$, so that all terms need only sum to zero at the origin, whereas state orthogonality ensures $F_{1,n}^\ast(0)=0$, in which case each contribution must vanish separately.

An interesting feature shown in the top- and bottom-left panels of Fig.~\ref{figF1} is that all contributions (in each dissection) to the $F_1^\ast$ form factor become comparable at large photon's momenta, and thus insights on the quark-diquark dynamical structure of nucleon resonances should be expected from the electro-production studies at high $Q^2$ which will be performed in the near future using the CLAS12 detector.

Regarding comparison with experiment, $F_{1,p}^\ast(x)$ agrees quantitatively in magnitude and trend with the data on $x\gtrsim 2$, an outcome which owes fundamentally to the QCD-derived momentum-dependence of the propagators and vertices employed in solving the bound-state and scattering problems. The mismatch on $x\lesssim 2$ between data and the prediction is also revealing. As we have emphasized, our calculation yields only those form factor contributions generated by a rigorously-defined dressed-quark core whereas meson-cloud contributions are expected to be important on $x\lesssim 2$. Thus, the difference between the prediction and data may plausibly be attributed to MB\,FSIs, as described in Sec.~5 of Ref.~\cite{Roberts:2016dnb}. (See also Refs.~\cite{Mokeev:2015lda, Kamano:2018sfb}).

\begin{figure}[!t]
\begin{center}
\begin{tabular}{cc}
\includegraphics[clip,width=0.43\linewidth]{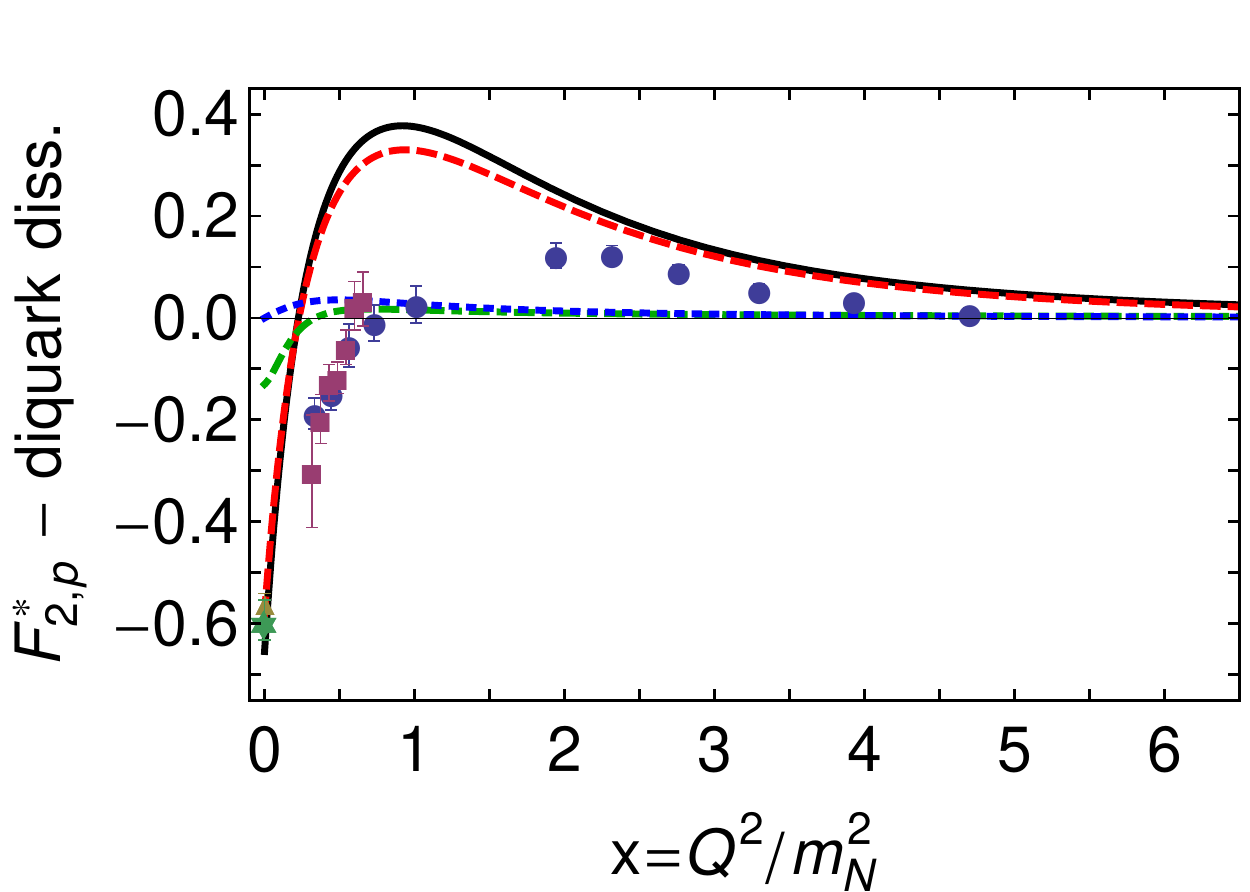} &
\includegraphics[clip,width=0.43\linewidth]{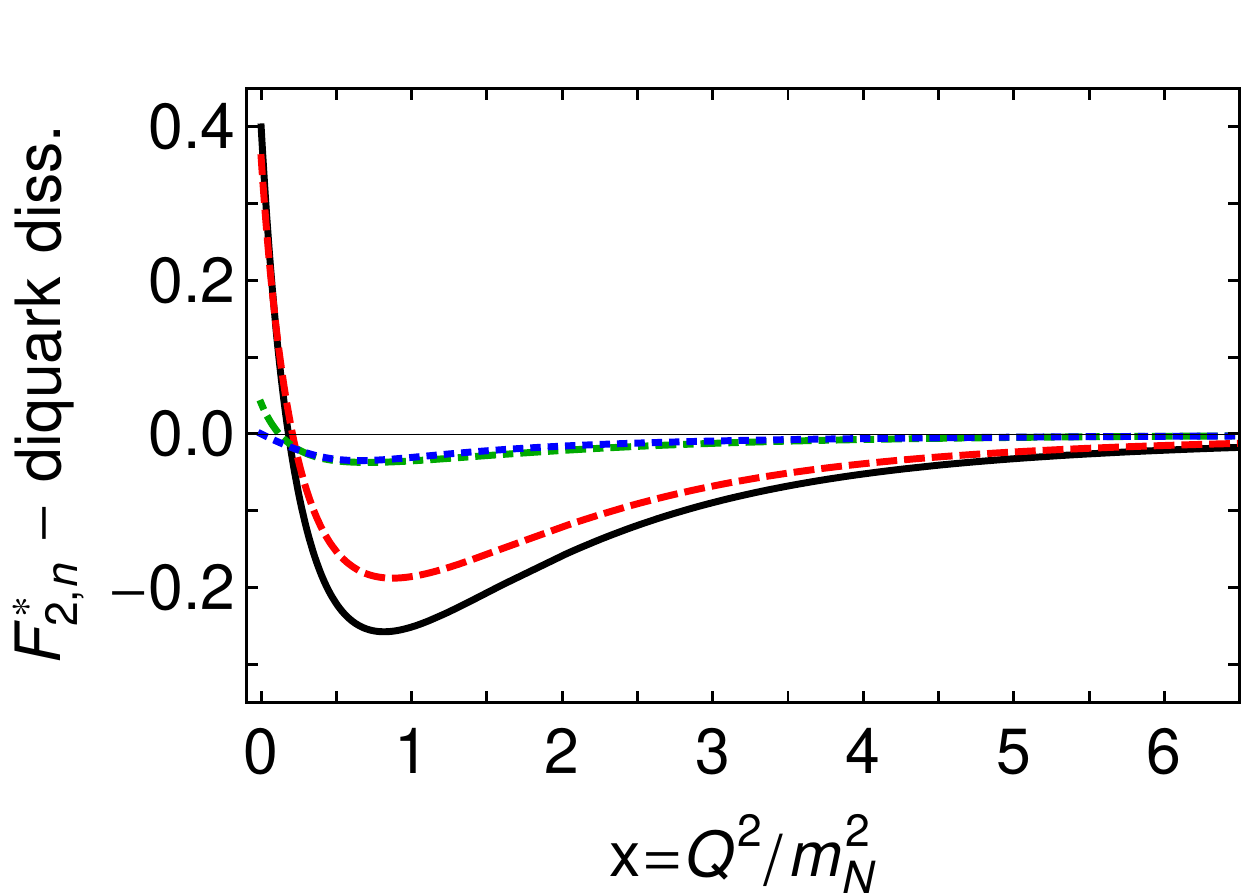}
\end{tabular}
\begin{tabular}{cc}
\includegraphics[clip,width=0.43\linewidth]{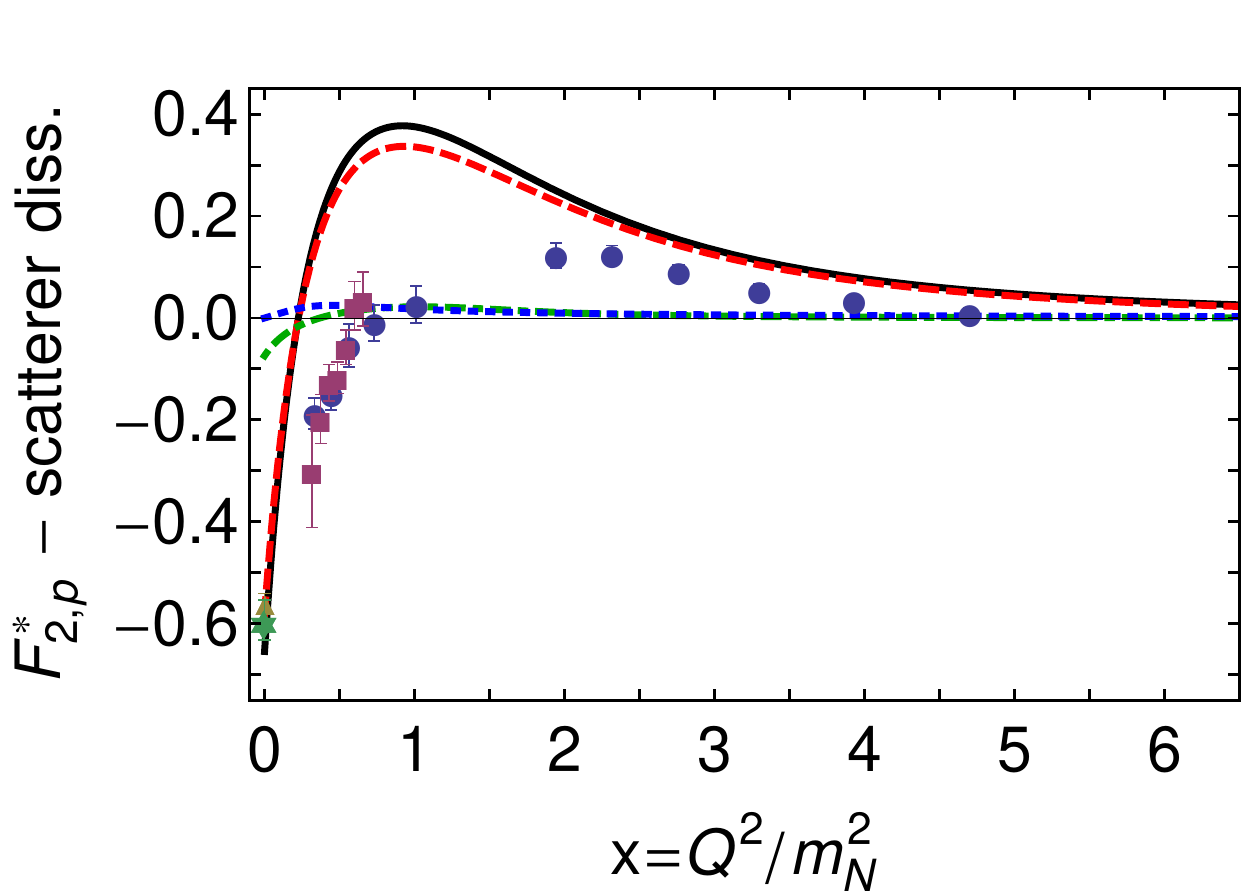} &
\includegraphics[clip,width=0.43\linewidth]{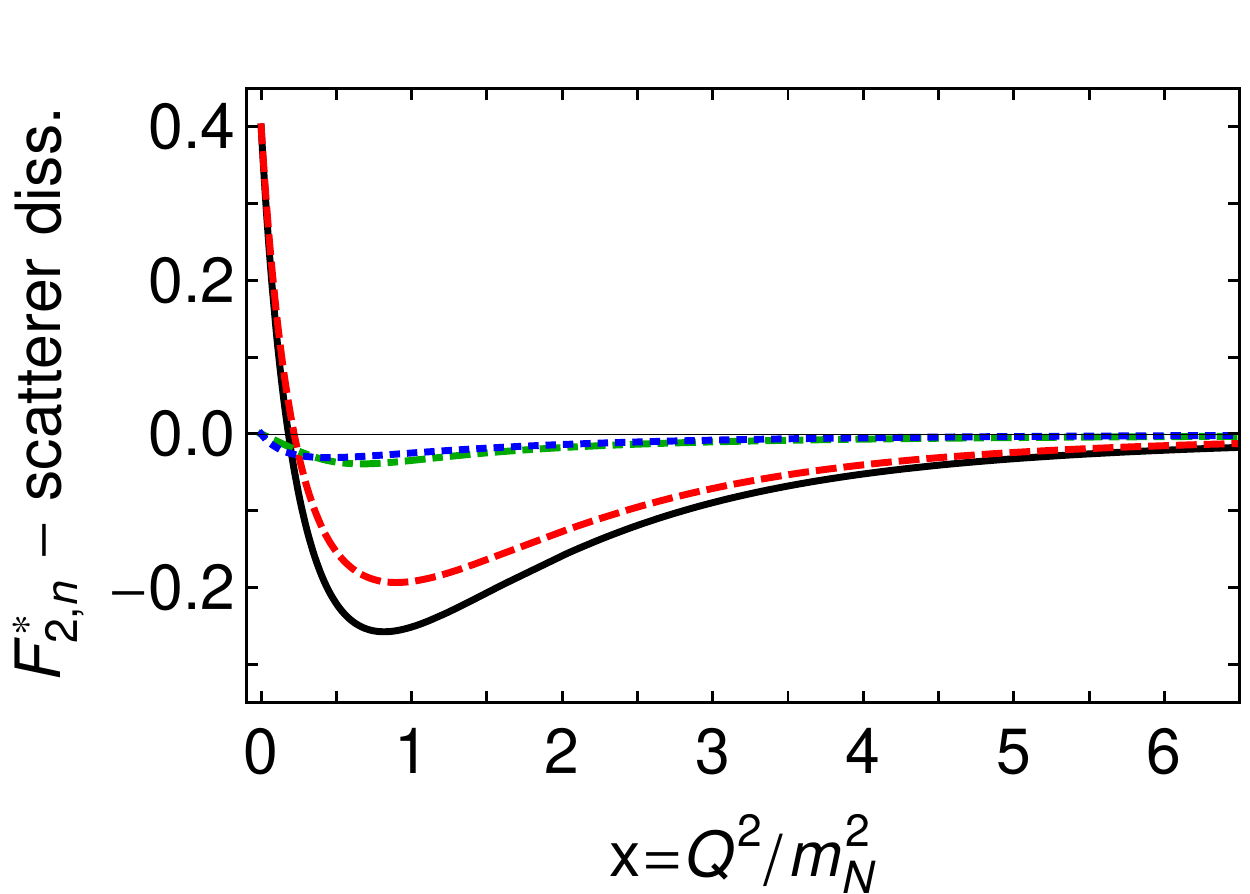}
\end{tabular}
\end{center}
\caption{\label{figF2} Computed Pauli transition form factor, $F_{2}^{\ast}$, for the charged reaction $\gamma^\ast\,p\to R^+$ (left panels) and the neutral reaction $\gamma^\ast\,n\to R^0$ (right panels): solid (black) curve in each panel.
Data:
circles (blue)~\cite{Aznauryan:2009mx},
squares (purple)~\cite{Mokeev:2012vsa, Mokeev:2015lda},
triangle (gold)~\cite{Dugger:2009pn},
and star (green)~\cite{Tanabashi:2018oca}.
\emph{Upper panels} -- diquark breakdown: \emph{DD1} (dashed red), scalar diquark in both nucleon and Roper; \emph{DD2} (dot-dashed green), pseudovector diquark in both nucleon and Roper; \emph{DD3} (dotted blue), scalar diquark in nucleon, pseudovector diquark in Roper, and vice versa.
\emph{Lower panels} -- scatterer breakdown: \emph{DS1} (red dashed), photon strikes an uncorrelated dressed-quark; \emph{DS2} (dot-dashed green), photon strikes a diquark; and \emph{DS3} (dotted blue), diquark breakup contributions, including photon striking exchanged dressed-quark.
\vspace*{-0.50cm}
}
\end{figure}

Pauli transition form factors for the $\gamma^\ast N \to R$ reaction are shown in Fig.~\ref{figF2}. They are all nonzero at $x=0$ and each possesses a zero crossing at roughly the same location, \emph{viz}.\, $x\approx 0.2$. Notably, as with $F_2^{p,R^+}$ and $F_2^{n,R^0}$ in Fig.\,\ref{elastic}, $F_{2,p}^\ast$ and $F_{2,n}^\ast$ are similar in magnitude and $Q^2$-dependence. In particular, the value of $F_{2,p}^\ast(0)/F_{2,n}^\ast(0) \approx -3/2$ is consistent with available data \cite{Tanabashi:2018oca}.

The remarks above concerning MB\,FSIs also apply to $F_2^\ast$; and, importantly, although they affect its precise location, the existence of a zero in $F_{2}^{\ast}$ is not influenced by MB\,FSIs. We are thus confident of our prediction for a zero in $F_2^{n,R^0}$. This zero will be found near that of $F_2^{p,R^+}$ if MB\,FSIs are not too different between these channels; and there are good reasons to suppose they are comparable because the two reactions are isospin-exchange partners and isospin symmetry is a good approximation for strong interactions.

\begin{figure}[!t]
\includegraphics[clip,width=0.45\linewidth]{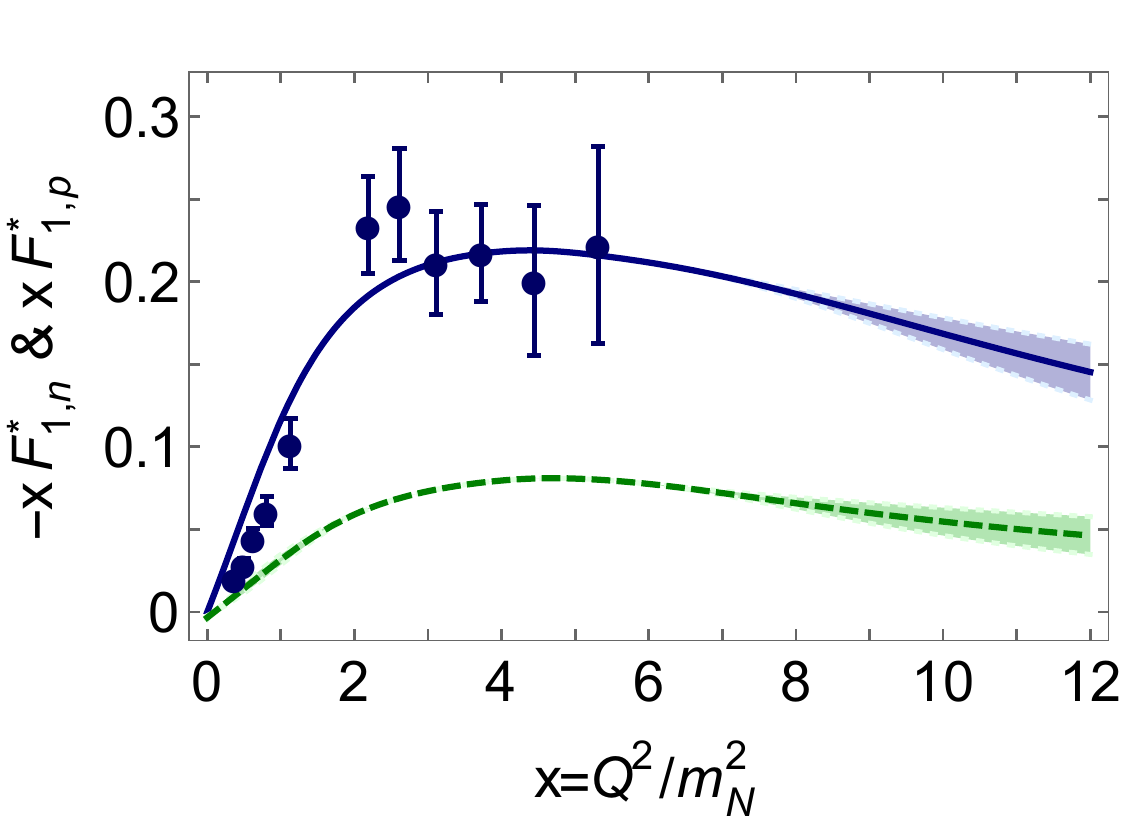}
\hspace*{0.50cm}
\includegraphics[clip,width=0.45\linewidth]{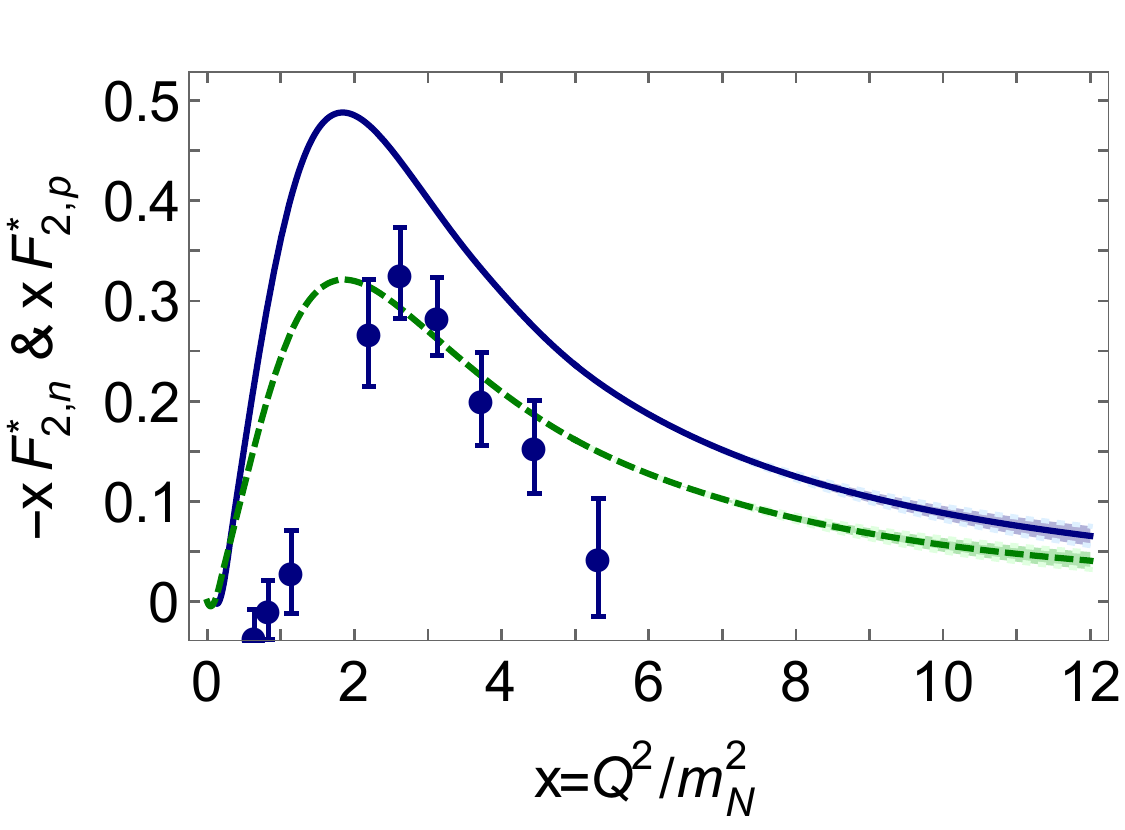}
\caption{\label{figLargeQ2}
Computed $x$-weighted Dirac (left panel) and Pauli (right panel) transition form factors for the reactions $\gamma^\ast\,p\to R^+$ (solid blue curves) and $\gamma^\ast\,n\to R^0$ (dashed green curves). In all cases, the results on $x\in [6,12]$ are projections, obtained via extrapolation of analytic approximations to our results on $x\in [0,6]$ (see Ref.~\cite{Chen:2018nsg} for details). The width of the band associated with a given curve indicates our confidence in the extrapolated value.
Data in both panels are for the charged channel transitions, $F_{1,p}^\ast$ and $F_{2,p}^\ast$: circles (blue)~\cite{Aznauryan:2009mx}. No data currently exist for the neutral channel.
\vspace*{-0.50cm}
}
\end{figure}

Finally, since it is anticipated that JLab~12 detector will deliver data on the Roper-resonance electroproduction form factors out to $Q^2 \sim 12 m_N^2$, we depict in Fig.\,\ref{figLargeQ2} the $x$-weighted Dirac and Pauli transition form factors for the reactions $\gamma^\ast p \to R^{+}$, $\gamma^\ast n\to R^{0}$ on the domain $0<x<12$. The results on $x>6$ are determined via the Schlessinger point method (SPM), as described in Ref.~\cite{Chen:2018nsg}. On the domain depicted, there is no indication of the scaling behaviour expected of the transition form factors: $F^\ast_{1} \sim 1/x^2$, $F^\ast_2 \sim 1/x^3$.  Since each dressed-quark in the baryons must roughly share the impulse momentum, $Q$, we expect that such behaviour will only become evident on $x\gtrsim 20$.

\vspace*{-0.50cm}
\section{Summary}
\label{sec:summary}

We have computed a range of properties related with the dressed-quark core of the proton's ground and first excited states using a Poincar\'e-covariant continuum approach to the three valence-quark bound-state problem in quantum field theory.

Amongst the results we have described, the following are of particular interest: (i) there are nonpointlike, fully-interacting quark-quark (diquark) correlations within these states and, in general, inside any kindred baryon; (ii) the isoscalar-scalar and isovector-pseudovector diquarks are dominant in these baryons and both, ground- and excited-state, have the same diquark relative content; (iii) the rest-frame wave functions of both states are largely $S$-wave in nature and the first excited state in this $1/2^+$ channel has the appearance of a radial excitation of the ground state; (iv) assuming that the first excited state of the nucleon is the so-called Roper resonance, we compare with experiment our computation of the equivalent Dirac and Pauli form factors of the $\gamma^\ast p \to R^+$ reaction and observe that, while the mismatch in the domain of $Q^2\lesssim 2 m_N^2$ may plausibly be attributed to meson-cloud effects, the agreement on $Q^2 \gtrsim 2 m_N^2$ owes fundamentally to the QCD-derived momentum-dependence of the propagators and vertices employed in solving the bound-state and scattering problems.

Let us finish this manuscript highlighting that novel experiments are approved at JLab\,12, and others are either planned or under consideration as part of an international effort to measure transition electro-couplings at large photon virtualities of all prominent nucleon resonances~\cite{Aznauryan:2012ba, Mokeev:2016hqv, Carman:2016hlp}. Therefore, our predictions herein but also forthcoming analyses that will involve low-lying negative-parity baryons and excited states of the $\Delta$-resonance will be thoroughly tested in the foreseeable future, and such efforts have the potential to deliver empirical information that would address a wide range of issues, including, \emph{e.g}.: is there an environment sensitivity of DCSB; and are quark-quark correlations an essential element in the structure of all baryons?


\vspace*{-0.20cm}
\begin{acknowledgements}
The material described in this contribution is drawn from work completed in collaboration with numerous excellent people, to all of whom I am greatly indebted.
I would also like to thank V. Mokeev, R. Gothe, V. Burkert and T.-S. H. Lee for insightful comments and support along the last years;
and to express my gratitude to the editors of the journal Few-Body Systems for inviting me to participate with an article in the {\it Ludwig Faddeev Memorial Issue}.
\end{acknowledgements}

\vspace*{-0.70cm}



%
%
%

\end{document}